\documentclass[useAMS,usenatbib]{mn2e}
\usepackage{amssymb}
\usepackage{amsmath}
\usepackage{graphicx}
\usepackage{rotating}
\usepackage{lscape}
\usepackage{array}
\def\Ha{H$\sf \alpha$}
\def\hi{H\,{\sc i}}
 
\def\deg{$^{\circ}$}
\def\kms{km~s$^{-1}$}
\def\msun{M$_{\odot}$}
\long\def\symbolfootnote[#1]#2{\begingroup%
\def\thefootnote{\fnsymbol{footnote}}\footnote[#1]{#2}\endgroup} 

\def\aj{AJ}%
\def\araa{ARA\&A}%
\def\apj{ApJ}%
\def\apjl{ApJ}%
\def\apjs{ApJS}%
\def\aap{A\&A}%
\def\mnras{MNRAS}%
\def\nat{Nature}%
\title[Three-dimensional modeling of the \hi\ kinematics of NGC~2915]{Three-dimensional modeling of the \hi\ kinematics of NGC~2915}
\author[Elson et al.]{E. C. Elson$^{1,2}$\thanks{Previous address: South African Astronomical Observatory, PO Box 9, Observatory 7935, South Africa}\thanks{E-mail:
ed.elson@icrar.org (ECE), edeblok@ast.uct.ac.za (WJGdeB), kraan@ast.uct.ac.za (RCK-K)}, W. J. G. de Blok$^{1}$ and R. C. Kraan-Korteweg$^{1}$\\
$^{1}$Astrophysics, Cosmology and Gravity Centre (ACGC), Department of Astronomy, University of Cape Town,\\ Private Bag X3,
Rondebosch 7701, South Africa\\
$^{2}$International Centre for Radio Astronomy Research, The University of Western Australia, M468,\\ 35 Stirling Highway, Crawley, WA, 6009 Australia\\}

\begin{document}

\date{Accepted 2011 March 15.}

\pagerange{\pageref{firstpage}--\pageref{lastpage}} \pubyear{2002}

\maketitle

\label{firstpage}
\begin{abstract}
The nearby blue compact dwarf, NGC~2915, has its stellar disc embedded in a large, extended ($\sim 22$~$B$-band scale-lengths) \hi\ disc.  New high-resolution \hi\ synthesis observations of NGC~2915 have been obtained with the Australia Telescope Compact Array.  These observations provide evidence of extremely complex \hi\ kinematics within the immediate vicinity of the galaxy's star-forming core.  We identify and quantify double-peaked \hi\ line profiles near the centre of the galaxy and show that the \hi\ energetics can be accounted for by the mechanical energy output of the central high-mass stellar population within time-scales of $10^6-10^7$~yr.  Full three-dimensional models of the \hi\ data cube are generated and compared to the observations to test various physical scenarios associated with the high-mass star-forming core of NGC~2915.  Purely circular \hi\ kinematics are ruled out together with the possibility of a high-velocity-dispersion inter-stellar medium at inner radii.  Radial velocities of $\sim 30$~\kms\ are required to describe the central-most \hi\ kinematics of the system.  Our results lend themselves to the simple physical scenario in which the young stellar core of the galaxy expels the gas outwards from the centre of the disc, thereby creating a central \hi\ under-density.  These kinematics should be thought of as being linked to a central \hi\ outflow rather than a large-scale galactic blow-out or wind.  

\end{abstract}


\section{Introduction}\label{intro}
Quantitatively understanding galaxy evolution is one of the major challenges facing modern astrophysics.  Star formation (SF) activity, along with merger activity and other environmental dependencies, largely dictates the evolutionary path of a galaxy.  Stellar feedback plays a particularly significant role in determining the distribution and kinematics of a galaxy's inter-stellar medium (ISM) on local and global length-scales.  In the so-called ``galactic fountain scenario'' \citep{galactic_fountain,bregmen_1980}, the ISM is heated and pushed out of the galaxy by stellar winds and supernova explosions.   The ejected material moves into the halo of the galaxy, expands, cools and falls back onto the disc.  Within the gaseous disc, stellar winds can dictate the ISM kinematics by setting up expanding gas shells and holes, typically with diameters of $\lesssim0.5$~kpc \citep[e.g.][for the cases of IC~2574; M31; M101 and NGC~6946 respectively]{IC2574,brinks_bajaja_1986,kamphuis_phd}.   In the case that the ejected gas is accelerated to velocities larger than the escape velocity of the galaxy, significant fractions of the ISM will escape into the inter-galactic medium (IGM) \citep[][]{maclow_1999}.  In this sense, stellar winds play an important role in the chemical evolution of galaxies and the IGM.  

\begin{figure}
	\begin{centering}
	\includegraphics[angle=0, width=1\columnwidth]{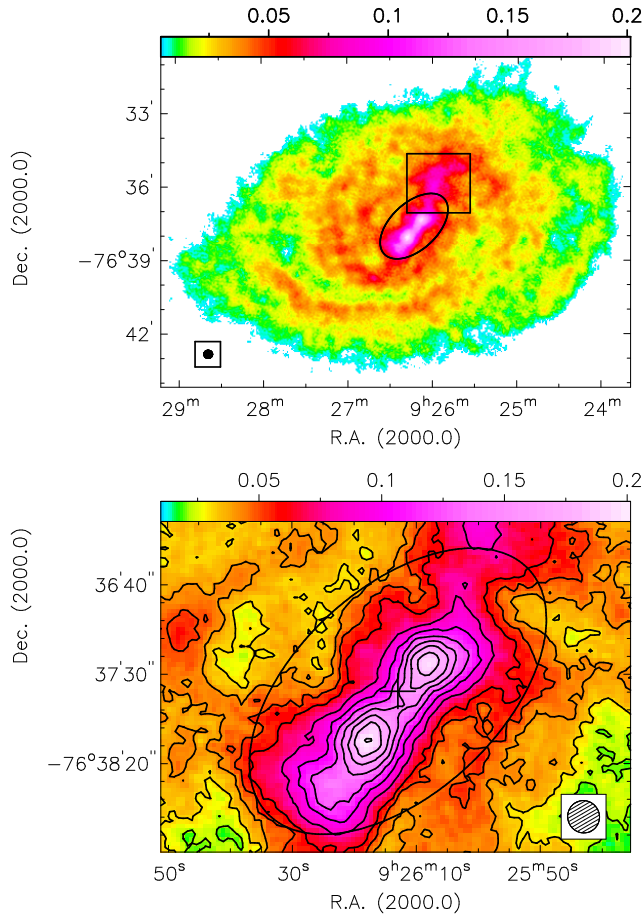}
	\caption{\hi\ total-intensity map of NGC~2915 (top panel) with a zoom-in of the central region (bottom panel), taken from  \citet{elson_2010a_temp}.  Contour levels run from  0.08~--~0.23~mJy~beam$^{-1}$ in steps of 0.02~mJy~beam$^{-1}$.  The intensity scale of each panel is specified in units of Jy~beam$^{-1}$.  The photometric centre determined by \citet{elson_2010a_temp} is marked with a cross.  The hatched circles in the lower corners represent the half power beam width of the synthesised beam.  The ellipses have a major axis length of 98~arcsec, and represent the $R$-band $R_{25}$ radius.  The rectangle delimits the plume-like \hi\ feature mentioned in Section~\ref{double_profiles}.}
	\label{slice_positions}
	\end{centering}
\end{figure}

Dwarf galaxies generally have shallow gravitational potential wells, meaning that the input of kinetic energy into the ISM from high-mass star formation and supernova explosions can significantly affect the gas dynamics.  With their typically low rotational shear rates and gas surface densities, they serve as useful laboratories to study the long-lived effects of stellar feedback processes.  In this paper we investigate the \hi\ dynamics of the nearby \cite[$\sim4.1\pm0.3$~Mpc,][]{meurer2003} blue compact dwarf NGC~2915 using new \hi\ synthesis observations obtained with the Australia Telescope Compact Array (ATCA).   \citet{meurer1} showed the optical appearance of NGC~2915 to be dominated by two main stellar populations: a blue compact core and a more diffuse, older red population.  The blue core is the locus of ongoing high-mass star formation.  This galaxy's seemingly typical nature, as suggested by its optical appearance, is contrasted by the presence of a large, extended ($\sim 22$~$B$-band scale-lengths) \hi\ disc with well-defined spiral structure \citep{meurer2,elson_2010a_temp}.  The \hi\ total intensity map from \citet{elson_2010a_temp} is reproduced in Fig.~\ref{slice_positions} with an ellipse that delimits the edge of the stellar disc.  NGC~2915 is also extremely dark-matter-dominated \citep{meurer2} with a total mass to blue light ratio as high as $M_{tot}/L_B\sim 140$~\msun/$L_B$ at the edge of its \hi\ disc \citep{elson_2010a_temp}.

Both \citet{meurer2} and \citet{elson_2010a_temp} provide evidence of complex \hi\ kinematics near the centre of the galaxy.  The sharp kinks and wiggles seen in the iso-velocity contours of their \hi\ velocity fields are indicative of non-circular velocity components within the gas.  \citet{2915_flows} found indirect kinematic evidence for an axisymmetric radial outflow of $5-17$~\kms\ near the centre of the \hi\ disc.  Direct evidence of broad and multi-component line profiles is provided by position-velocity slices extracted from the central regions of the \hi\ disc of NGC~2915 \citep{elson_2010a_temp}.  These non-Gaussian line profiles are in the immediate vicinity of the galaxy's star-forming core, and could therefore be due to the effects on the ISM of stellar winds and supernova explosions associated with the young stellar population.  In this paper we present the results of a detailed study of the kinematics of the central \hi\ characteristics of the galaxy, with the intention of identifying and quantifying the underlying physical processes that lead to the observed \hi\ properties.  A simple physical scenario that links the effects of high-mass star formation at the centre of the galaxy to the observed \hi\ kinematics is proposed and tested. 

The layout of this paper is as follows.  The \hi-line data set is presented in Section~\ref{data}.  Double-peaked \hi\ line profiles are identified and quantified in Section~\ref{double_profiles}, with a corresponding discussion of the \hi\ energetics appearing in Section~\ref{central_energetics}.  Various three-dimensional models of the \hi-line data are compared to the observations and discussed in Section~\ref{models}.  Based on the observed \hi\ kinematics, the possible fate of the central \hi\ component is discussed in Section~\ref{fate}.  Our general conclusions are presented in Section~\ref{conclusions}.
\section{\hi\ data}\label{data}
In this work we utilise the \hi-line data of \citet{elson_2010a_temp}.  Using six different antenna configurations of the ATCA, nearly 100 hours worth of on-source observations were obtained for NGC~2915 between 2006 October~23 and 2007 June~2 (project number C1629).  \citet{elson_2010a_temp} describe the various telescope setups of the observations as well as the standard data reduction procedures carried out to arrive at the calibrated, continuum-subtracted, deconvolved \hi\ data cubes.  This analysis makes use of their naturally-weighted \hi\ data cube which has an angular resolution of 17.0~arcsec~$\times$~18.2~arcsec and a channel spacing of 3.2~\kms.  The noise in a line-free channel is Gaussian-distributed with a standard deviation of $\sigma\sim 0.6$~mJy~beam$^{-1}$.
\section{Double-peaked \hi\ line profiles}\label{double_profiles}
\begin{figure}
	\begin{centering}
	\includegraphics[angle=0,width=1\columnwidth]{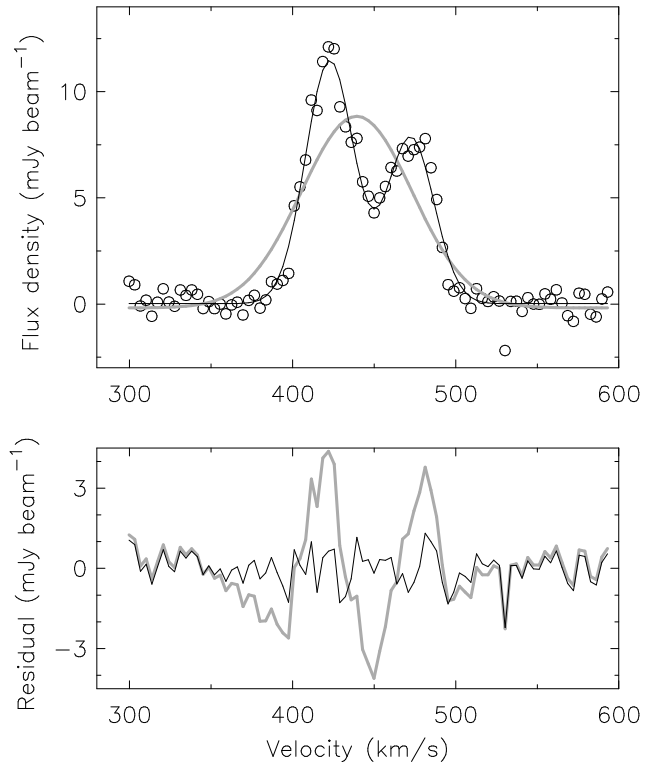}
	\caption{Example of a double-peaked line profile extracted from position $\alpha_{2000}$~=~09$^\mathrm{h}$~26$^\mathrm{m}$~12.6$^\mathrm{s}$, $\delta_{2000}$~=~$-76$\deg~37$'$~37.8$''$ in the re-gridded \hi\ data cube of NGC~2915.  This position is marked by a black cross in Fig.~\ref{chi2_map}.  In the top panel, open circles represent the data, the solid grey and black curves represent best-fitting single- and double-component Gaussians, respectively.  The solid grey and black curves in the lower panel represent the residuals for the fitted single- and double-component Gaussians, respectively.}
	\label{double_peaked}
	\end{centering}
\end{figure}

\citet{elson_2010a_temp} demonstrated the existence of multi-component and broad \hi\ line profiles within NGC~2915's \hi\ disc.  Many of the profiles within $\sim 150$~arcsec of the centre of the galaxy are observed to be double-peaked.  In order to classify a particular line profile as being either single- or double-component in nature, the reduced $\chi^2$ goodness-of-fit statistics of fitted single- and double-component Gaussians were compared.  A line profile with a double Gaussian goodness-of-fit statistic, $\chi^2_d$, that was less than 85~per~cent of the corresponding single Gaussian goodness-of-fit statistic, $\chi^2_s$ (i.e.~$\chi^2_d<0.85\chi^2_s$) was classified as being double-peaked.  Profiles for which $\chi^2_d>0.85\chi^2_s$ were classified as consisting of a single component.  The \hi\ data cube was re-gridded to a pixel size of 10~arcsec~$\times$~10~arcsec in order to increase the level of independence between adjacent line profiles.  Three filters were used simultaneously when fitting the line profiles with single- and double-component Gaussians: (1) line profiles with a fitted peak less than 4$\sigma$ above the noise were rejected; (2) line profiles with a fitted line-width less than the channel width were rejected; and (3) fitted profile peaks had to be within the velocity range of the data cube.  An example of a fitted double-peaked profile is shown in Fig.~\ref{double_peaked}.  
   
\begin{figure}
	\begin{centering}
	\includegraphics[angle=0,width=1\columnwidth]{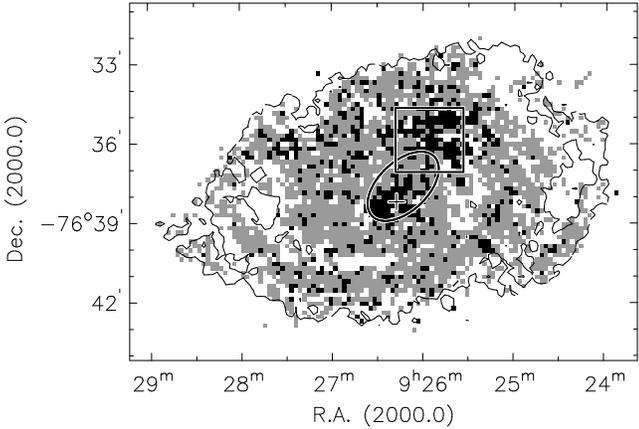}
	\caption{Comparison of the quality of single- and double-component Gaussian fits to the \hi\ line profiles.  Black-filled squares denote the positions at which the reduced $\chi^2$ statistic of the fitted double Gaussian is less than 85~per~cent of the corresponding statistic for the fitted single Gaussian, i.e. $\chi^2_d<0.85\chi^2_s$.  Grey-filled squares represent profiles  for which $\chi^2_d>0.85\chi^2_s$.  White-filled squares represent line profiles that were rejected by the fitting filters and which were therefore not parameterised.  The angular size of each coloured square is 10~arcsec~$\times$~10~arcsec, corresponding to $\approx$~183~pc~$\times$~183~pc.  The single contour is at an \hi\ column density level of $1.6\times 10^{20}$~cm$^{-2}$, and is arbitrarily selected to represent the approximate edge of the outer \hi\ disc.  The ellipse represents the edge of the old stellar disc and has a semi-major axis length of 98~arcsec.  The rectangle is the same as the one shown in Fig.~\ref{slice_positions} and delimits the plume-like \hi\ feature seen in the \hi\ total intensity map.  The cross at $\alpha_{2000}$~=~09$^\mathrm{h}$~26$^\mathrm{m}$~12.6$^\mathrm{s}$, $\delta_{2000}$~=~$-76$\deg~37$'$~37.8$''$ marks the position of the \hi\ profile shown in Fig.~\ref{double_peaked}.}
	\label{chi2_map}
	\end{centering}
\end{figure}

A comparison between the reduced $\chi^2$ goodness-of-fit statistics of both sets of fits showed that of the 2571 fitted profiles, 561 have $\chi^2_d<0.85\chi^2_s$, i.e. they are significantly better fit by a double-component Gaussian than by a single-component Gaussian.  These profiles are shown as black-filled squares in Fig.~\ref{chi2_map} while the profiles for which $\chi^2_d>0.85\chi^2_s$ appear in grey.  It is evident that the double-peaked \hi\ profiles are most prevalent within the immediate vicinity of the galaxy's star-forming core.  Double-peaked profiles also occur within the general vicinity of the plume-like \hi\ feature seen in the \hi\ total intensity map (delimited by a rectangle in Figs.~\ref{slice_positions} and \ref{chi2_map}).  \citet{elson_2010a_temp} show this feature within the \hi\ distribution to correspond to asymmetric \hi\ line profiles on the receding side of the galaxy.  In position-velocity slices these \hi\ profiles appear as an  `\hi\ beard' that is lagging in velocity relative to the main disc.  The authors estimate the mass of the plume to be $\sim 5.6$~per~cent of the total \hi\ mass.  Although kinematically interesting, the plume-like \hi\ feature is not considered any further in this work.  Also evident from Fig.~\ref{chi2_map} is the fact that most of the double-peaked line profiles near the centre of the galaxy are located close to the south-eastern central \hi\ concentration seen in the \hi\ total intensity map.  For each of the identified double-peaked \hi\ line profiles within 98~arcsec of the dynamical centre (i.e. within the black ellipse shown in Fig.~\ref{chi2_map}), the absolute difference between the two fitted peak velocities was determined.  The average velocity separation for the profiles is $\Delta V=59$~\kms, roughly 72~per~cent of the galaxy's asymptotic circular velocity of $81.9\pm 1.6$~\kms\ \citep{elson_2010a_temp}.  

\section{Central energetics}\label{central_energetics}
Are the effects of stellar winds and/or supernovae explosions the cause of the double-peaked \hi\ line profiles observed near the centre of NGC~2915?  Several authors have linked the effects of stellar feedback to radially expanding gas components in nearby galaxies.  For example, by carrying out a kinematic study of the irregular dwarf galaxy NGC~4816 using \hi\ and H$\alpha$ observations, \citet{van_eymeren} detected prominent \hi\ and H$\alpha$ outflows with expansion velocities of $V_{exp}\sim$~25~\kms\ and $V_{exp}\sim$~30~\kms, respectively.  They related these outflows to a fast-expanding supergiant gas shell, set up by the injection of mechanical energy from massive stars into the ISM.  \citet{IC2574}, in their study of the holes and shells in the ISM of the dwarf galaxy IC 2574, determined the expansion velocity of the gas around each of 40 identified HII regions.  They found expansion velocities of $V_{exp}\sim 6 - 25$~\kms.  One of their main findings was that the energy requirements for ``the formation of the holes can be understood in terms of the combined effects of stellar winds and multiple supernova explosions of the most massive stars formed during a recent phase of active star formation''.  \citet{young_et_al_2003} used VLA observations of three dwarf galaxies to analyse their \hi\ line profile widths and shapes.  The authors found a trend between the fraction of asymmetric profiles in a galaxy and its \Ha\ luminosity, with galaxies having greater star formation rates also having a greater fraction of asymmetric, double-peaked line profiles.  They interpret the result as being indicative of the star formation activity and its associated injection of kinetic energy into the ISM that is responsible for stirring the surrounding \hi.  

\subsection{Gas energetics}
Supposing that the double-peaked \hi\ line profiles observed in the central regions of NGC~2915 are due to an expanding gas component, could the young stellar core be supplying the energy required to drive such an expansion?  To begin answering this question, an estimate of the energy associated with the expanding gas is required.  Following \citet{meurer2}, the central energetics are quantified by estimating the kinetic energy of an expanding gas component as
\begin{equation}
E_k={1\over 8}M_g\Delta V^2,
\end{equation}
where $ M_g$ is the mass of the expanding gas and $ \Delta V$ is the observed velocity separation between the peaks of a double-peaked \hi\ line profile.  The observed velocity separation is  related to the expansion velocity by $ V_{exp}={1\over 2}\Delta V$.  For the sake of brevity, the term ``splitting'' is henceforth used to denote the velocity separation between the fitted peaks of a double-peaked \hi\ line profile.  Considering only the split line profiles within 98~arcsec of the centre of NGC~2915 (i.e. black-filled resolution elements within the ellipse shown in Fig.~\ref{chi2_map}), the corresponding flux density of each profile (obtained from the corresponding resolution element in the re-gridded \hi\ total intensity map) was converted to a mass estimate, all of which were summed to yield $ M_g\sim 2.6\times 10^6$~\msun\ for the mass of expanding gas near the galaxy centre.  The average splitting of $\Delta V=59$~\kms\ for the central double-peaked line profiles suggests an expansion velocity of $V_{exp}\sim30$~\kms.  These $ M_g$ and $ \Delta V$ values yield $ E_k\sim5.6\times10^{44}$~J for the amount of kinetic energy that can be linked to an expanding central gas component. 

\citet{meurer2}, using $ M_g=10^8$~\msun\ and $ \Delta V=40$~\kms, estimated $E_k\sim4\times 10^{46}$~J for NGC~2915 assuming a distance of 5.1~Mpc.  At a distance of 4.1~Mpc, their energy estimate reduces to $E_k\sim2.6\times 10^{46}$~J.  It is stressed that our estimate of $ M_g$ is a lower limit for the mass of expanding gas.  Only the masses corresponding to identified split profiles were considered, as opposed to all of the mass within a specified central region.  For this reason, our $E_k$ estimate is lower than that of \citet{meurer2}.  The expanding gas mass $M_g=10^8$~\msun\ used by \citet{meurer2} to estimate $E_k$ is almost as large as the total \hi\ mass of NGC~2915 derived by \citet{elson_2010a_temp} from the global \hi\ profile of their high-resolution \hi\ data cube.  We therefore treat the $E_k\sim2.6\times 10^{46}$~J estimate of \citet{meurer2} as an upper limit for the kinetic energy of the expanding neutral ISM.

\subsection{Stellar energetics}\label{stellar_energetics}
\begin{figure}
	\begin{centering}
	\includegraphics[angle=0,width=1\columnwidth]{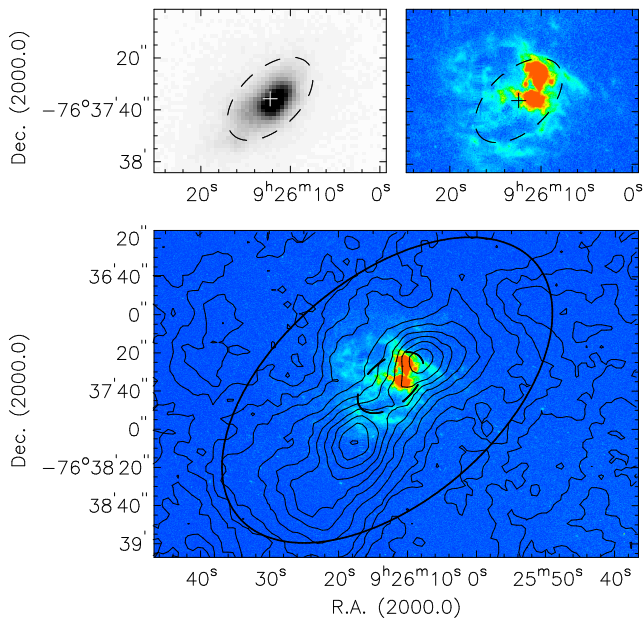}
	\caption{Top left panel: {\sc galex} near ultra-violet image of NGC~2915.  Top right panel: H$\alpha$ image of NGC~2915 produced using the data from \citet{gil_de_paz_et_al_2003}.  Bottom panel: \Ha\ image with \hi\ flux density contours overlaid.  \hi\ contour levels run from 10~-~60~mJy~beam$^{-1}$ in steps of 10~mJy~beam$^{-1}$ and from 60~-~230~mJy~beam$^{-1}$ in steps of 20~mJy~beam$^{-1}$.  The \Ha\ intensity scale is un-calibrated in the top right and bottom panels. In all panels the dashed black ellipses, included to facilitate length-scale comparisons between the images, have a semi-major axis of length 20~arcsec.   The solid black ellipse in the bottom panel is the same ellipse shown in Figs.~\ref{slice_positions}~and~\ref{chi2_map}, and represents the approximate edge of the old stellar disc.  The crosses in the top panels represent the photometric centre as determined by \citet{elson_2010a_temp}.}
	\label{halpha}
	\end{centering}
\end{figure}

The amount of kinetic energy linked to the expanding gas component needs to be compared to the mechanical energy output of the central stellar population. \Ha\ emission, a tracer of recent and ongoing high-mass star formation, is observed near the centre of NGC~2915.  Importantly, it \emph{only} occurs at inner radii where a large fraction of the split \hi\ line profiles are observed.  An \Ha\ image produced using the data of \citet{gil_de_paz_et_al_2003} is shown in the top-right panel of Fig.~\ref{halpha}.  The {\sc galex} near ultra-violet image is shown in the top-left panel.  The location of the \Ha\ emission relative to the \hi\ emission is shown in the bottom panel.  The \Ha\ emission is clearly contained between the two central \hi\ concentrations seen in the \hi\ total intensity map.  The filamentary structure of the \Ha\ emission is indeed testament to the perturbing effects of the high-mass star formation on the distribution of the ISM.  As an example, the large shell-like structure seen in the south-east region of the H$\alpha$ distribution is spatially coincident with the split \hi\ line profiles near the centre of the galaxy.  


While the intense ultra-violet radiation energy produced by young, massive O- and B-type stars is capable of ionising the neutral ISM, it is not expected to play a significant role in redistributing it.  Rather, it is the mechanical energy injected by stars via stellar winds and supernova explosions that is expected to set up shock-heated (T~$\gtrsim$~10$^6$~K) gas and expanding structures within the ISM.  \citet{chu_2005} neatly demonstrates that the total amount of mechanical wind energy output during the entire lifetime of an O-type star, including the high-mass-loss-rate Wolf-Rayet phase, is $\sim 4 \times 10^{43}$~J.  The author further points out that this energy is more than a factor of two lower than the mechanical explosion energy of a supernova, $\sim 10^{44}$~J.  In this work we therefore ignore the energy deposited into the ISM by stellar winds and consider only the mechanical energy produced by supernova explosions.  In the remainder of this section we estimate the time-scales over which the mechanical energy deposition from supernovae into the ISM can account for the $E_k\sim5.6\times10^{44}$~J and $E_k\sim2.6\times10^{46}$~J estimates presented above for the ISM kinetic energy. 

\citet{leitherer_heckman} show that the equilibrium ratio of the mechanical luminosity from supernova explosions to a galaxy's \Ha\ luminosity is $L_{mech}^{SNe}/L_{H\alpha}=1.8$.  This, they say, is the case for a Salpeter IMF \citep{salpeter_IMF} stellar population having formed at a constant rate.  Using this relation together with the $L_{H\alpha}~=~10^{32.5}$~J~s$^{-1}$ measurement of \citet{gil_de_paz_et_al_2003} yields an estimate of $L_{mech}^{SNe}~=~5.7\times 10^{32}$~J~s$^{-1}$ for the total mechanical luminosity of NGC~2915's supernova population.  However, not all of this mechanical energy will be deposited into the ISM as kinetic energy.  For example, \citet{mori_ferrara_madau_2002} treat only 30~per~cent of the mechanical energy from supernovae as being converted into ISM kinetic energy.  Adopting the same conversion factor, we estimate $L_{mech}^{ISM}~=~0.3 L_{mech}^{SNe}~=~1.7\times10^{32}$~J~s$^{-1}$ to be the rate at which mechanical energy from supernovae is injected into NGC~2915's ISM.  At this rate the central \hi\ energetics estimates of $E_k\sim 5.6\times10^{44}$ to $2.6\times 10^{46}$~J can be accounted for within $\sim 1.0\times 10^5$ to $4.8\times 10^6$~yr.  

A second estimate for the timescale required by NGC~2915's supernova population to energise the central HI component of the galaxy can be obtained by considering the expected total supernova mass and supernova rate.  Treating a single supernova event as depositing $0.3\times10^{44}$~J of mechanical energy into the ISM, roughly 20~-~860 supernovae are needed to account for amounts $E_k\sim 5.6\times10^{44}$~-~$2.6\times 10^{46}$~J of kinetic energy associated with the expanding central \hi\ component.  As a galaxy produces stars that will later explode as supernovae, it builds up stellar mass.  A question therefore is whether NGC~2915's observed stellar mass is consistent with the galaxy having formed 20~-~860 type-II supernova progenitors over its stellar assembly lifetime.  To answer this question we randomly sampled stellar masses between 0.08~\msun\ and 120~\msun\ from a Kroupa IMF \citep{kroupa_IMF}.  As the masses were sampled, supernova progenitors were identified as those stars more massive than 8~\msun.  860 supernovae were found to be accounted for by a total stellar mass of only $\sim 6.7\times 10^4$~\msun.  These supernovae constituted $\sim 26$~per~cent of the total stellar mass, with an average supernova mass of $\sim19.8$~\msun.  The supernova masses were distributed as $M_{SN}^{-2.12\pm0.10}$.  The remaining 74~per~cent of the total stellar mass was hence made up by stars less massive 8~\msun.  A total stellar mass of $6.7\times 10^4$~\msun\ is much less than that of the blue stellar population of NGC~2915.  Adopting a $B$-band extinction-corrected apparent magnitude of $m_B=12.0$ \citep{ngc2915_mB,schlegel} together with the 4.1~Mpc distance estimate of \citet{meurer2003}, the total $B$-band luminosity of NGC~2915 is determined to be $3.2\times 10^8$~L$_{\odot}$.  Even for a $B$-band stellar mass-to-light ratio as low as 0.1~\msun/L$_{\odot}$, the total mass of the blue stellar population is still at least two orders of magnitude larger than that required for $\sim 860$ type-II supernovae to form.

The supernova rate of a galaxy is often found to be of the order of one per~cent of the total star formation rate  \citep{boomsma_thesis}.  \citet{kennicutt_1983} determined the relation
\begin{equation}
 SFR_{tot}=3.42\times10^{-8}L_{H_\alpha},
\end{equation}
where $ SFR_{tot}$ is the total star formation rate of a galaxy in units of \msun\ yr$^{-1}$ and $ L_{H\alpha}$ its total \Ha\ luminosity in solar units.  Using this relation together with the $L_{H\alpha}~=~10^{32.5}$~J~s$^{-1}$ measurement of \citet{gil_de_paz_et_al_2003} yields an estimate of $SFR_{tot}=2.8\times 10^{-2}$ \msun~yr$^{-1}$ for the total star formation rate of NGC~2915.  The supernova rate is thus expected to be of the order of 10$^{-4}$~\msun~yr$^{-1}$.  Considering only type-II supernovae events, a lower supernova mass limit for 860 supernovae is $M_{SN}=860\times8\approx6.8\times 10^3$~\msun, since core collapse supernovae events only occur for stars at least as massive as 8~\msun.  The ratio of this supernova mass to the supernova rate yields a formation timescale of $\sim 6.8\times 10^7$~yr.  The corresponding timescale for 20 supernovae is $\sim1.6$~Myr.

These estimated time-scales of $\sim10^5 - 10^7$~yr match the typical main sequence lifetimes of  high-mass O- or B-type stars.  These are exactly the types of stars expected to constitute NGC~2915's blue stellar core, thereby implicating them as a mechanical energy source responsible for the anomalous central \hi\ kinematics.  In the sections that follow, the central \hi\ component of NGC~2915 is modelled as being radially expanding due to the input of kinetic energy from the high-mass stars into the ISM.  The kinematic models that we produce are not expected to precisely match the complex features of the \hi\ data.  Instead, they allow us to generally identify the constituent disc properties and processes that may be responsible for the observed \hi\ distribution and kinematics.
\section{Three-dimensional models}\label{models}
A radially expanding inner gas component should exhibit distinct kinematic signatures as seen in a three-dimensional data cube.  To check whether such a scenario is consistent with the observed \hi\ kinematics of NGC~2915, various model cubes were constructed and compared to the observations.  The model cubes differ in the radial profiles of some of their orientation parameters as well as their kinematics.  

\subsection{Modeling routine}
The model cubes were generated using the task {\sc halomodgal}\symbolfootnote[3]{Kindly provided by Filippo Fraternali.}, a modified version of the standard {\sc gipsy} task {\sc galmod}.  The routine assumes axisymmetry and distributes many ``\hi\ clouds'' into a set of tilted rings.  For each ring the inclination, position angle, and rotation velocity must be provided; as well as a velocity dispersion, column density and scale-height perpendicular to the plane.  Unlike the standard {\sc galmod} routine, the user can additionally specify non-circular velocities parallel and perpendicular to the plane for each ring using {\sc halomodgal}.  The routine selects random positions at which to distribute \hi\ clouds within each ring.  Having placed a cloud, the routine uses the systemic and rotation velocities of the tilted ring to build a velocity profile around the position of the cloud.  For this purpose the cloud is divided into sub-clouds and each sub-cloud assigned a velocity equal to the central velocity of the cloud\symbolfootnote[4]{The sum of the cloud's systemic and rotation velocities.} plus a random increment extracted from a Gaussian.  This Gaussian has a mean equal to the central velocity and a width equal to the velocity dispersion of the ring.  Finally, the routine uses the position and velocity of the sub-cloud to place it at the corresponding pixel position within the data cube.

\subsection{\hi\ disc scale-height}\label{scale_height}
The {\sc halomodgal} routine requires an estimate of the \hi\ disc scale-height.  To obtain an estimate, the method outlined by \citet{IC2574} was used.  They state that for a density distribution given by $ \rho(z,R)=2 \rho(0,R)$~$\mathrm{sech}(z/z_0)$ \citep{kellman_1972,van_der_kruit_1981}, the scale-height is
\begin{equation}
{ z_0(R)={\sigma_{gas}\over \sqrt{2\pi G\rho(0,R)}}},
\label{z0}
\end{equation}
where $ \sigma_{gas}$ is the velocity dispersion of the gas in quiescent regions of the galaxy and $ \rho(0,R)$ is the total volume density of the galaxy in the disc plane.  Under this approximation, the \hi\ disc scale-height is proportional to the \hi\ velocity dispersion and inversely proportional to the total mass density in the disc.  Assuming a Gaussian distribution of the \hi\ disc in the $z$-direction, the $1\sigma$ scale-height is approximately $ h=z_0/\sqrt{2}$.

A constant value of 10~\kms\ was used for $ \sigma_{gas}$, chosen in accordance with the typical azimuthally-averaged second-order moments of the \hi\ line profiles of the outer \hi\ disc of NGC~2915 \citep{elson_2010a_temp}.  Following \citet{IC2574} to calculate the average mass density in the plane of the gas disc, it was assumed that the total mass of the galaxy can be estimated from the last measured point (with radius $ R_{last}$) of the observed rotation curve, $ V_{last}$.  This mass was treated as being distributed over a spherical volume of radius $ R_{last}$.  The results of the tilted ring models fitted to the \hi\ velocity field by \citet{elson_2010a_temp} were used: $ R_{last}=8.7$~kpc and $ V_{last}=86$~\kms.  These values yield an average density of $\rho=0.005$ \msun~pc$^{-3}$ at the radius $ R_{last}$.  Since this is only the mean density of the galaxy, and the density near the plane is certainly higher, \citet{IC2574} suggest that twice the derived value should be used to approximate the mid-plane density.  The mid-plane density is thus approximated as $ \rho(0,R)=0.01$~\msun~pc$^{-3}$.  Under a minimum disc assumption, \citet{elson_2010a_temp} parameterised the dark matter halo of NGC~2915 as a pseudo-isothermal sphere.  They estimated a central dark matter density of $\rho_0=0.17\pm0.03$~\msun~pc$^{-3}$ and a core radius of $R_c=0.9\pm0.1$~kpc.  Our estimate for the mid-plane dark matter density is therefore roughly a tenth of the core density, and matches the halo volume density at a galactocentric radius of 3.6~kpc.  Using these $ \sigma_{gas}$ and $ \rho(0,R)$ estimates together with Eqn. \ref{z0} and $ h=z_0/\sqrt{2}$ yields $h=413$~pc as the approximate scale-height of the gaseous disc NGC~2915.  This thick \hi\ layer is attributed to the lower gravitational potential of the system (as compared to high-surface-brightness late-type spirals) and the \hi\ velocity dispersion of $\sigma_{gas}\sim 10$~\kms.

\subsection{Tested models}

\begin{figure*}
	\begin{centering}
	\includegraphics[angle=0, width=2\columnwidth]{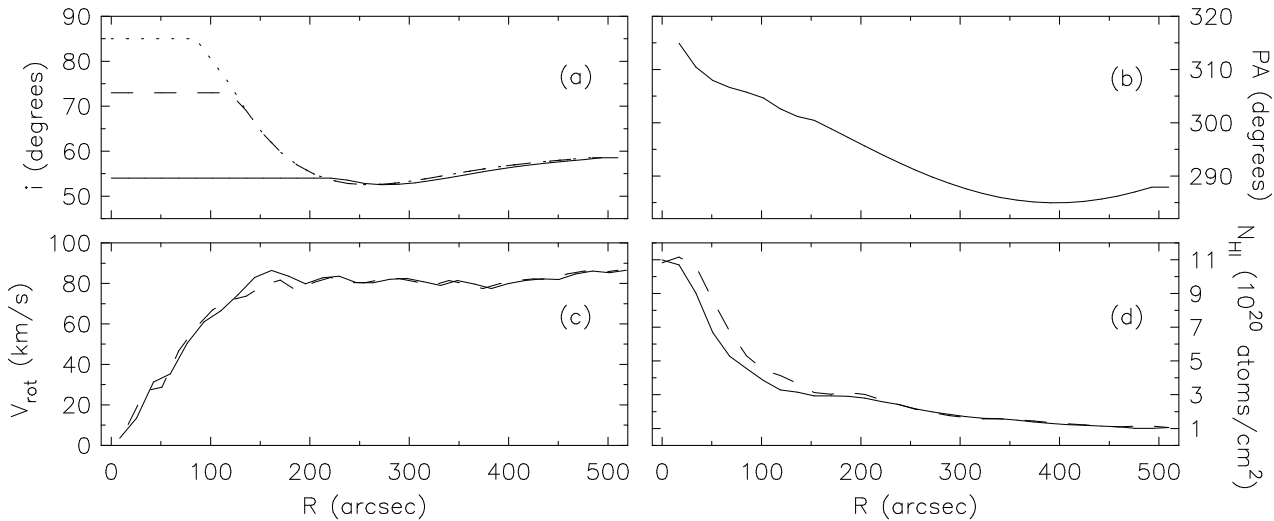}
	\caption{Radial profiles used to construct the model cubes.  Panels (a), (b), (c) and (d) represent inclination, position angle, circular velocity and face-on column density radial profiles, respectively.  Solid and dashed curves represent the profiles from tilted ring models CI and SI of \citet{elson_2010a_temp} respectively.  Note that these tilted ring models have the same position angle radial profile, shown as a single curve in panel (b).  The dotted curve in panel (a) represents the inclination radial profile used to construct model cube~SI+exp (Sec.~\ref{radial_flow_models}).}
	\label{panel_plot}
	\end{centering}
\end{figure*}

Several models were generated to study the \hi\ kinematics of NGC~2915.  To create these models, the tilted ring modeling results of \citet{elson_2010a_temp} were used.  Specifically, the inclination and position angle radial profiles as well as the rotation curves of their so-called tilted ring models CI and SI were incorporated.  These profiles are shown in Fig.~\ref{panel_plot}a, b and c.  The two tilted ring models differ mainly in their inclination profiles, and very slightly in their rotation curves.    Model~CI has an almost constant inclination of $i\sim 55$\deg\ for all radii.  The inclination profile of model~SI rises from $i\sim$~55\deg\ to $i\sim$~75\deg\ when moving from the outer to the inner disc, beginning at a galactocentric radius of $\sim 240$~arcsec.  For all model cubes, the systemic velocity was always kept constant at 465 \kms.  The kinematic centre was set equal to the photometric centre as determined by \citet{elson_2010a_temp}.  The \hi\ column density profile from \citet{elson_2010a_temp} was inclination-corrected using each of the inclination radial profiles, thereby yielding the two face-on column density radial profiles shown in Fig.~\ref{panel_plot}d.  Each of the model cubes was smoothed down to a resolution of 17.0~arcsec~$\times$~18.2~arcsec, the resolution of the actual data cube.  

Various position-velocity slices were extracted from the model cubes and compared directly to the corresponding slices from the \hi\ data cube.  The locations and orientations of these slices in the \hi\ total intensity map are shown in the bottom-most panel of Fig.~\ref{panel_slices}.  Each position-velocity slice was chosen to investigate a particular component or feature of the complex \hi\ kinematics of NGC~2915.  Slice~1 targets the outer disc and is used to demonstrate its relatively simple circular kinematics.  Slice~2 is placed along the kinematic minor axis of the \hi\ disc in order to probe radial and other non-circular gas motions.  The general bar-like morphology of the inner \hi\ disc, including the two \hi\ concentrations near the galaxy's centre, is studied by means of slice~3.  Finally, slice~4 is selected to further highlight the distinct non-circular motions near the centre of the \hi\ disc.   

\subsubsection{Purely circular models}
Six different model cubes were constructed in total.  All of the model position-velocity data are shown in Fig.~\ref{panel_slices} together with the observed data.  The first two are purely circular models, and were made using the radial parameter profiles of tilted ring models CI and SI from \citet{elson_2010a_temp}, together with a constant \hi\ velocity dispersion of 10~\kms.  These two model cubes, henceforth referred to as cubes CI and SI, served as the ``base'' models.  They were manipulated in order to construct the rest of the model cubes.  Cubes CI and SI are shown in columns 1 and 2 of Fig.~\ref{panel_slices}.  Slice 1 (row 1) demonstrates how these purely circular models are capable of reproducing the observed \hi\ kinematics of the outer disc where non-circular gas motions are negligible.  Slice 1 also shows that, by construction, all six models reduce to models CI and SI in the outer disc.  Slices 2, 3 and 4, however, show that simple, purely circular models such as cubes CI and SI are unable to reproduce the inner-disc \hi\ kinematics of NGC~2915 which are dominated by non-circular velocity components.

\subsubsection{Variable dispersion models}
Is a neutral ISM with a high velocity dispersion responsible for the non-Gaussian line profiles near the centre of the galaxy?  The second-order \hi\ moment map presented by \citet{elson_2010a_temp} exhibits a sharp increase from outer to inner disc, with second-order moments as high as $\sim 25$~\kms\ near the centre of the galaxy.  Is this sharp increase in the second-order moments of the \hi\ line profiles an accurate representation of the \hi\ velocity dispersions, or is it an artefact of the split line profiles near the centre of the galaxy?  To check whether such an increase in velocity dispersion is consistent with the \hi\ data, the second-order moment map radial profile from \citet{elson_2010a_temp} was added as a velocity dispersion radial profile to model cubes CI and SI.  The resulting model cubes are henceforth referred to as cubes CI+disp and SI+disp, and are presented in columns 3 and 4 of Fig.~\ref{panel_slices}.  Panels 10, 11, 12, 14, 15, 16 of Fig.~\ref{panel_slices} show that although these dispersion models are able to reproduce the large spreads in central velocities, they cannot match the double-peaked structure of the \hi\ line profiles.  A central ISM with a high velocity dispersion therefore cannot explain the central gas kinematics of NGC~2915.  

\subsubsection{Radial flow models}\label{radial_flow_models}
\begin{figure*}
	\begin{centering}
	\includegraphics[angle=-90, width=2\columnwidth]{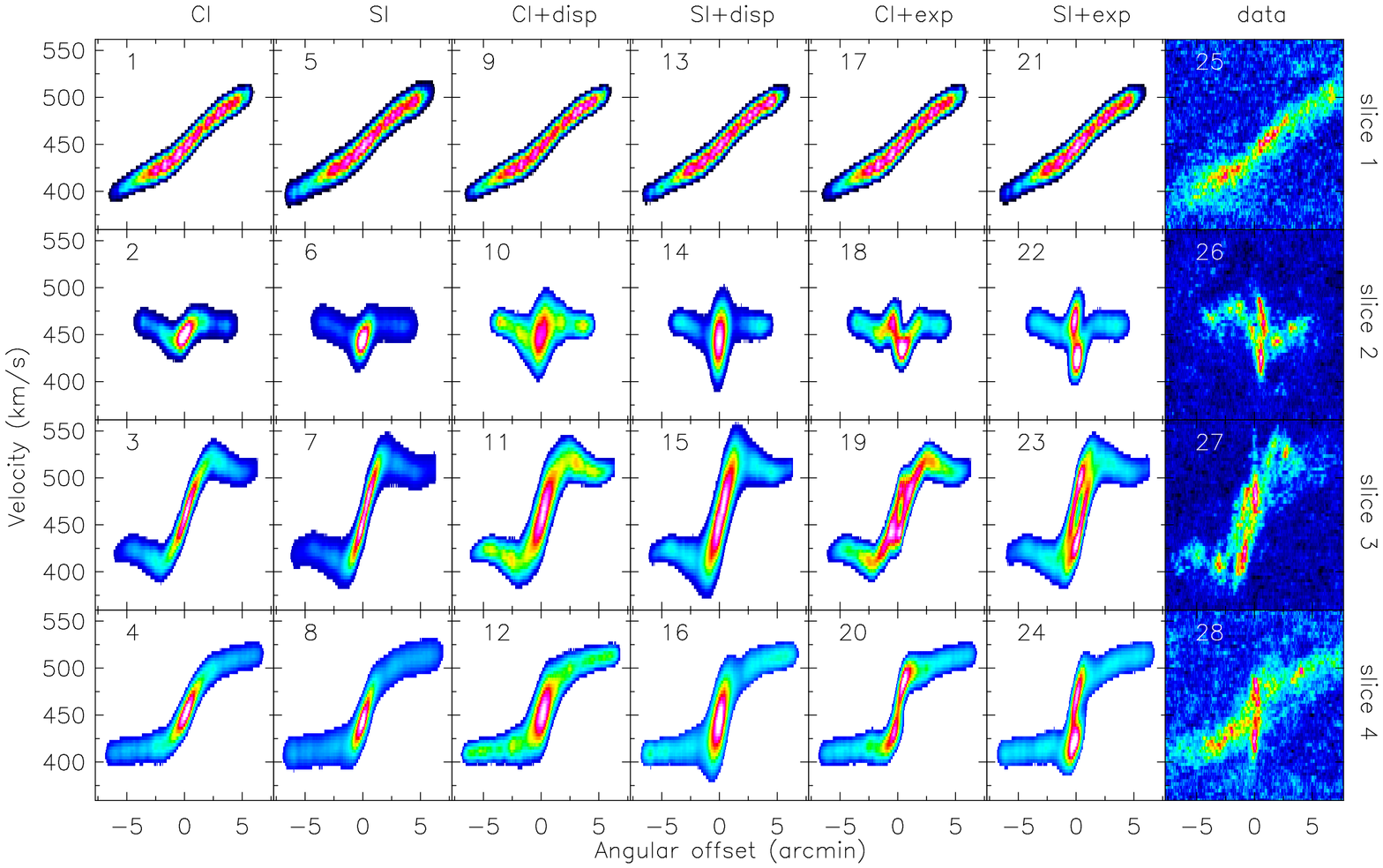}
	\includegraphics[angle=0, width=\columnwidth]{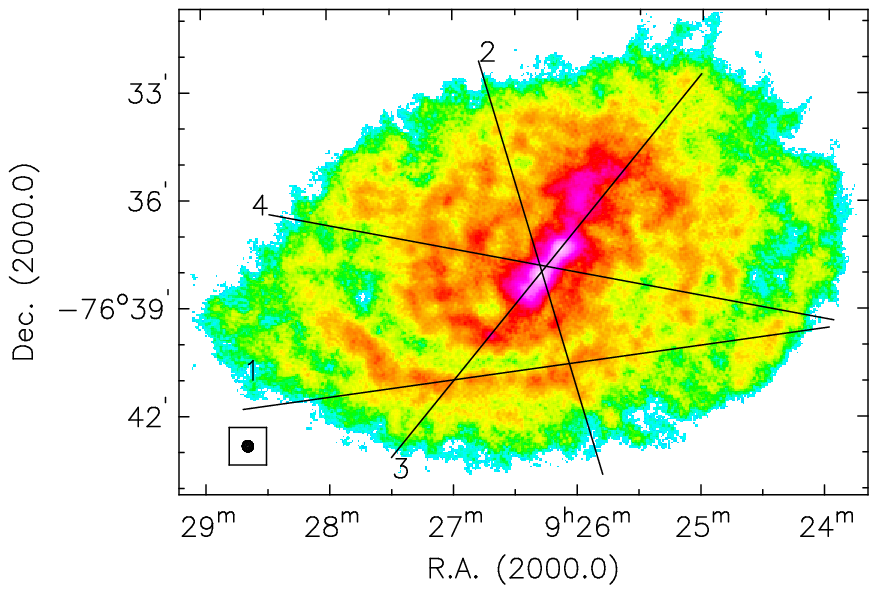}
	\caption{Top panels: Position-velocity slices extracted from the model and \hi\ data cubes of NGC~2915.  Each row represents a different position-velocity slice, with the slice locations and orientations in the \hi\ total intensity map shown in the lower panel.  \newline \textbf{Comments on columns:} Each of columns 1-6 (from left to right) represent a different model cube of a thin, rotating gas disc.  Columns 1, 2: constant velocity dispersion of 10~\kms\ and no expansion velocities; columns 3, 4: high velocity dispersion towards the centre and no expansion velocities; columns 5, 6: constant velocity dispersion of 10~\kms\ and expansion velocities of $\sim$~30~\kms\ near the centre.  Column 7 shows the \hi\ data cube position-velocity slices.}
	\label{panel_slices}
	\end{centering}
\end{figure*}

Since none of the purely circular models were able to match the \hi\ kinematics of the inner disc, the final set of models included radial velocities.  The results of Sec.~\ref{double_profiles} above show that, on average, \hi\ line profiles are split by $\sim 60$~\kms\ within the central $\sim 100$~arcsec of the galaxy.  If this observed line splitting is interpreted  as being due to an expanding gas component, the expansion speed is equal to half of the observed line splitting.  Model cubes CI+exp and SI+exp were therefore constructed by adding radial expansion velocities of $V_{exp}=30$ \kms\ at all radii $R\le 85$~arcsec to model cubes CI and SI.  These models are shown in columns 5 and 6 of Fig.~\ref{panel_slices}.  A distinct trend in the general structure of the model cube line profiles is noticeable when moving from column 1 to column 6 (left to right) in Fig.~\ref{panel_slices}: columns 1 and 2 have single-component, relatively low-dispersion Gaussian-shaped line profiles.  The remaining columns 3~-~6 have line profiles with higher second-order moments, yet columns 3 and 4 still only contain single-component line profiles.  Most noticeable in columns 5 and 6 is the double-peaked nature of the inner-most line profiles, with the more specific nature of the splitting differing between columns 5 and 6.

The split line profiles seen in the model cubes (panels 18, 19, 20, 22, 23, 24) are mostly consistent with the double-peaked profiles seen in the data (panels 26, 27, 28).  Clearly, however, it is cube SI+exp that best matches the data.  The only significant difference between cubes CI+exp and SI+exp is the inclination of the inner disc, which was found to play a crucial role in determining the double-peaked nature of the central line profiles.  Various inner disc inclinations in the range 70\deg~-~80\deg, as suggested by the tilted ring models from \citet{elson_2010a_temp}, were experimented with.  All such inclinations were found to result in relatively poor matches to the observations in terms of split line profiles.  Model SI+exp was finally constructed using a constant inner disc inclination of 85\deg\ out to a galactocentric radius of 140~arcsec.  Figure~\ref{panel_plot}a shows the inclination radial profile used for cube SI+exp (dotted curve) together with the profiles used for the purely circular models.  Without this highly-inclined inner disc, the observed splitting of the line profiles at inner radii could not be matched.  Radial velocities of order $V_{exp}\sim$~30~\kms\ together with a highly-inclined inner disc are therefore crucial in terms of the description of the central gas kinematics of NGC~2915.

\subsection{Discussion}\label{discussion}
The distribution and kinematics of the \hi\ near the centre of NGC~2915 are consistent with the simple physical scenario in which the young stellar core of the galaxy, with which the \Ha\ emission is related, expels the gas outwards from the centre of the disc.  This process creates a central \hi\ under-density and a radially expanding central gas component.  Similar processes are observed in other nearby galaxies.  As mentioned in Sec.~\ref{central_energetics},   \citet{van_eymeren} detected prominent \hi\ and H$\alpha$ outflows in NGC~4816, with expansion velocities of $V_{exp}\sim$~25~\kms\ and $V_{exp}\sim$~30~\kms, respectively.  Such outflows can significantly affect the overall evolution of a galaxy.  Star-bursts, with their associated gas expulsion processes, are indeed thought to be a possible evolutionary link between late- and early-type dwarfs.  If a burst of star formation experienced by a gas-rich late-type dwarf is able to remove or at least redistribute a significant mass fraction of the ISM, star-formation could be quenched, resulting in the galaxy fading to become a red, gas-poor early-type dwarf.  The redistribution of gas within a galaxy can also lead to episodic star formation activity, with the removal and subsequent replenishment of \hi\ being linked to corresponding epochs of diminished and intensified star formation activity \citep[e.g.][]{williams_2010}.  Indeed, colour-magnitude diagrams of resolved stellar populations of nearby galaxies provide clear evidence of complex, episodic star formation histories \citep[e.g.][]{carina_project,m81_SFH}.

Although the high-resolution \hi\ observations together with our modeling results presented above suggest NGC~2915 to contain relatively little \hi\ mass at its centre, it could be the case that molecular hydrogen (H$_2$) fills the central region.  \citet{bigiel_2008} showed that a Schmidt-type power law with index $N = 1.0\pm 2$ relates the observed star formation rate surface density to the H$_2$ surface density across	a	sample	of	spiral	galaxies from The \hi\ Nearby Galaxy Survey \citep{THINGS_walter}, implying that H$_2$ forms stars at a constant efficiency. This relation allows the molecular gas content of a system to be estimated from a measure of its star formation rate surface density.  Based on the integrated star formation rate surface density of the  stellar core of NGC~2915 to be presented in our upcoming publication (Elson~et~al., 2011), a total molecular gas mass of $M_{H_2}=5.7^{+3.4}_{-2.1}
\times 10^7$~\msun\ is estimated to be co-located with the stellar disc of NGC~2915.  This amount of molecular gas is conceivably sufficient to fill the region between the central \hi\ concentrations.

Unclear is the reason why the inner portion of the \hi\ disc of NGC~2915 should be so highly-inclined relative to the outer portion.  It may be the case that the disc is severely warped.  Numerical simulations of dissipative mergers carried out by \citet{barnes_2002} have shown that extremely warped gas discs can result from the merging process.  \citet{barnes_2002} showed further that extremely warped gas disc merger remnants can result even in the case of unequal mass mergers.  A warped gas structure can also develop in the case that gas from the inter-galactic medium accretes onto the inner portions of a galaxy's gas disc \citep{bournaud_combes_2003}.  The degree of the warping will depend on the ratio of the angular momentum of the in-falling gas to that of the disc.  Although direct evidence for the above-mentioned scenarios is lacking in the case of NGC~2915, they cannot be ruled out as possible causes of a highly-warped inner \hi\ disc.  A good example of a galaxy with an extremely warped disc is the ``Spindle'' galaxy (NGC~2685).  Until recently, this system was generally regarded as being a two-ringed polar galaxy, consisting of two kinematically distinct discs aligned almost perpendicular to one another.  \citet{spindle_josh} recently showed, however, that NGC~2685 forms a coherent, extremely warped disc which creates the appearance of two mutually inclined regions.
\section{Fate of the expanding gas}\label{fate}
Having established that the central \hi\ kinematics of NGC~2915 are consistent with those of a radially-expanding gas component, the eventual fate of the expanding gas can be speculated on.  Will the central gas eventually stop expanding and fall back towards the centre, thereby fuelling a subsequent burst of star formation, or will it escape into the inter-galactic medium?  

Several authors have studied gas outflows in nearby dwarf galaxies (see \citealt{meurer_1705_1992} and \citealt{martin_IZW18_1996} for NGC~1705 and IZw~18 respectively).  In VVI Zw403, \citet{papaderos_1994} detected extended X-ray plumes which they interpreted as the outflow of hot gas.  \citet{maclow_1999} modelled the effects of repeated supernova explosions from starbursts in dwarf galaxies on the ISM, taking into account the gravitational potential of their dark matter halos.  They found that the mass ejection efficiency is very low in galaxies with a total mass $\gtrsim10^7$~\msun.  More specifically, they found that in galaxies more massive than $\sim~10^6$~\msun, the fraction of mass that is lost in a galactic blowout is as little as $10^{-6}$.  For reference, the dynamical mass for NGC~2915 is $\sim1.5\times10^{10}$~\msun\ \citep{elson_2010a_temp}.  Importantly, \citet{maclow_1999} showed that metal-enriched material from stellar winds and supernova ejecta is easily accelerated to velocities higher than the escape speed of the galaxy, especially in the case of low-mass systems.  This result, together with the fact that dwarf galaxies dominate the cosmic scenery in terms of number density  \citep[][]{springel_2005,sandage_virgo_LF,mateo_dwarfs_review_1998}, implies that dwarf galaxies play a crucial role in regulating the metallicity of the inter-galactic medium.

The gravitational potential of NGC~2915 is dominated by its dark matter halo.  Treating the density profile of the dark matter halo to be a pseudo iso-thermal sphere, the escape velocity of the system is given by
\begin{equation}
 V_{esc}(R)=\sqrt{2V_c^2\left(1+\ln\left[{R_{max}\over R}\right]\right)},
\label{vesc}
\end{equation}
where $ V_c$ is the asymptotic value of the circular rotation velocity at large radii, and $R_{max}$ is the maximum radius of the pseudo-isothermal sphere.  Based on the rotation curve derived for NGC~2915 by \citet{elson_2010a_temp}, we adopt $ V_c\sim86$~\kms\ and $ R_{max}\sim8.7$~kpc.  Using these values, an escape velocity radial profile was determined using Eqn.~\ref{vesc}.  For the sake of brevity, we simply mention that the escape velocities at all radii are $\gtrsim 175$ \kms, much larger than the inferred expansion velocities of $V_{exp}\sim 30$~\kms.  Furthermore, the dark matter halo of NGC~2915 is undoubtedly larger than the radial extent of the gas disc, implying that these escape velocities are lower limits.  The expanding gas in the centre of NGC~2915 therefore stands very little chance of escaping into the inter-galactic medium, and is consistent with a central outflow rather than a galactic blowout or wind.  Assuming an outflow time-scale of $\sim 100$~Myr, the average mass redistribution rate is $\sim 0.021$~\msun~yr$^{-1}$, roughly consistent with the estimated total star formation rate for the system, $SFR_{tot}\sim 0.028$~\msun~yr$^{-1}$ (Section~\ref{stellar_energetics}).  Despite being unable to escape the gravitational potential of the galaxy, the redistributed gas will significantly affect the future star-forming properties of NGC~2915's stellar disc.  
\section{Summary and conclusions}\label{conclusions}
We have used new deep, high-resolution \hi\ synthesis observations of NGC~2915 to carry out a detailed study of its \hi\ kinematics which are known to be extremely complex within the immediate vicinity of the young stellar core.  Sharp twists and kinks are seen in the iso-velocity contours of the \hi\ velocity field, and central \hi\ line profiles are known to be double-peaked and broad.  We parameterised the \hi\ line profiles, checking which of them are better described as a double-component Gaussian as opposed to a single-component Gaussian.  More than 20~per~cent of the \hi\ line profiles were found to be double-peaked, with many of these profiles co-located with the stellar disc of the galaxy.  The average line splitting was measured to be $\Delta V=59$~\kms, which if interpreted in terms of an expanding gas component implies expansions velocities of $\sim 30$~\kms\ for roughly $2.1\times 10^6$~\msun\ of \hi.

\Ha\ emission is observed to spatially coincide with the identified double-peaked \hi\ line profiles at inner radii.  Assuming the deposition of mechanical energy from supernovae into the ISM, the energetics of a possibly radially-expanding inner \hi\ component can be accounted for within $\sim 10^6 - 10^7$~yr, the typical main sequence lifetimes of high-mass O or B type stars.  The high-mass stellar disc near the centre of the galaxy is thus implicated as being the energising source responsible for the anomalous \hi\ kinematics.

The main focus of this paper has been the construction of full three-dimensional models of the NGC~2915 \hi\ data cube.  These models were used to test various physical scenarios which might be consistent with the system's observed \hi\ kinematics.  Models based on purely circular dynamics are able to reproduce the observed kinematics of the outer \hi\ disc, yet fail to accurately reproduce those of the inner disc.  Furthermore, the possibility of a high-velocity-dispersion inner portion of the \hi\ disc, which may ordinarily be related to high levels of star formation activity, is discredited.  These variable-dispersion models are able to reproduce the large spreads in observed central velocities, yet cannot match the double-peaked structure of the \hi\ line profiles.  A final set of models constructed by adding radial expansion velocities 30~\kms\ at radii spanning the spatial extent of the stellar disc are able to reproduce the main kinematic features of the inner \hi\ disc.  The optimal model consists of a highly-inclined ($i=85$\deg) inner \hi\ disc, without which the specific nature of the split \hi\ profiles could not be matched.  A highly-inclined inner disc could be the result of a dissipative merging process or perhaps a cold gas accretion event.  Although direct evidence for such scenarios is lacking in the case of NGC~2915, they cannot be excluded as possible causes of a highly-inclined inner disc.  Due to the extreme dark matter halo properties of NGC~2915, our identified expanding central gas component stands essentially no chance of escaping into the inter-galactic medium.  At all radii the escape velocities are a factor of $\sim 6$ larger than the estimated expansion velocities.  Assuming a time-scale of $\sim 100$~Myr, the central gas is being expelled outwards at a rate that is roughly consistent with the estimated total star formation rate.  The central \hi\ kinematics of NGC~2915 should therefore be thought of as forming a central outflow rather than a large-scale galactic blowout or wind.  
\section{Acknowledgements}\label{acknowledgements}
All authors thank Prof.~R.~Sancisi for extremely informative and fruitful discussions regarding the three-dimensional modeling of the \hi\ data cube.  The work of ECE was based upon research generously supported by the South African SKA project.  All authors acknowledge funding received from the South African National Research Foundation.  The work of WJGdeB is based upon research supported by the South African Research Chairs Initiative of the Department of Science and Technology and the National Research Foundation.  The Australia Telescope Compact Array is part of the Australia Telescope which is funded by the Commonwealth of Australia for operation as a National Facility managed by CSIRO.  Finally, all authors thank the anonymous referee for constructive comments that improved the quality of the paper.

\label{lastpage}

\end{document}